# Crowding Reveals Fundamental Differences in Local vs. Global Processing in Humans and Machines

Doerig, A.[†], Bornet, A.[†], Choung, O. H., Herzog, M. H.

Laboratory of Psychophysics, Brain Mind Institute, Ecole Polytechnique Fédérale de Lausanne (EPFL), Switzerland

[†] These authors contributed equally to this work.

## Abstract

Feedforward Convolutional Neural Networks (ffCNNs) have become state-of-the-art models both in computer vision and neuroscience. However, human-like performance of ffCNNs does not necessarily imply human-like computations. Previous studies have suggested that current ffCNNs do not make use of global shape information. However, it is currently unclear whether this reflects fundamental differences between ffCNN and human processing or is merely an artefact of how ffCNNs are trained. Here, we use visual crowding as a well-controlled, specific probe to test global shape computations. Our results provide evidence that ffCNNs cannot produce human-like global shape computations for principled architectural reasons. We lay out approaches that may address shortcomings of ffCNNs to provide better models of the human visual system.

## Introduction

Vision is a complex process that remained beyond the reach of computer systems for decades. Only recently, deep feedforward Convolutional Neural Networks (ffCNNs) have shown tremendous success in an impressive number of computer vision tasks, ranging from object recognition (Krizhevsky, Sutskever, & Hinton, 2012) and segmentation (Girshick, Radosavovic, Gkioxari, Dollár, & He, 2018), to image synthesis (Goodfellow et al., 2014; Karras, Laine, & Aila, 2018) and scene understanding (Eslami et al., 2018). ffCNNs and the human visual system share several similarities. For example, after training on complex visual datasets such as ImageNet (Deng et al., 2009), ffCNN neural activities show high correlations with human and non-human primate neural activities (Khaligh-Razavi & Kriegeskorte, 2014; Nayebi et al., 2018; Yamins et al., 2014) and the receptive fields of neurons in the earlier layers of these ffCNNs are qualitatively similar to those in the retina and early visual cortex (Lindsey, Ocko, Ganguli, & Deny, 2019; Zeiler & Fergus, 2014). Because of these similarities, ffCNNs trained on complex visual tasks were proposed as models of the human visual system (Khaligh-Razavi & Kriegeskorte, 2014; Kietzmann, McClure, & Kriegeskorte, 2018; Nayebi et al., 2018; VanRullen, 2017; Yamins et al., 2014). However, human-like performance of ffCNNs does not necessarily imply human-like computations. Importantly, several studies have shown that ffCNNs usually rely on local features while humans strongly rely on global shape information (Baker, Lu, Erlikhman, & Kellman, 2018; Brendel & Bethge, 2019; Doerig, Bornet, et al., 2019; Kim, Bair, & Pasupathy, 2019).

There are two main options to explain why ffCNNs do not process global shape like humans. First, this difference may come from *training*. ffCNNs are typically trained on ImageNet. It is interesting and surprising that local features seem to be the easiest way for these networks to classify natural images. However, a different training set in which local features are not predictive of the classes may require networks to rely on global shape computations. To address this possibility, Geirhos et al. (2018) created a new dataset in which textural information was of no avail for object recognition. They used a textural algorithm (Gatys, Ecker, & Bethge, 2016) to randomly swap textures in ImageNet. For example, the texture of a cat image was replaced by elephant-skin texture. This training dataset biased an ffCNN (ResNet50; He, Zhang, Ren, & Sun, 2016) towards shape-level features, because textural information was no longer useful for classifying this dataset. They validated the network's shape-bias by showing increased robustness to local noise and textural changes.

Alternatively, ffCNNs may be incapable of matching human global computations for principled *architectural* reasons. Even though Geirhos et al.'s network was able to ignore local features, it may not use global computations in the same way as humans. One difficulty in addressing this question is that there is no consensus about how to experimentally diagnose *how* deep networks compute global information.

To specifically investigate local vs. global processing in humans and machines, we use visual crowding as an experimental probe. Crowding is the technical term for the everyday observation that objects are harder to perceive in clutter. Neighbouring visual elements are perceived as jumbled or indistinct, and are hard to recognize (Fig. 1; reviews: Herzog, Sayim, Chicherov, & Manassi, 2015; Levi, 2011; Whitney & Levi, 2011). This phenomenon is strongest in the periphery, but also occurs in the fovea (Malania, Herzog, & Westheimer, 2007; Sayim, Westheimer, & Herzog, 2010) . This phenomenon is ubiquitous in natural vision since elements rarely appear in isolation (Fig. 1a). Crowding can also be studied with high precision in psychophysical experiments. For example, when a vernier target (i.e., two vertical bars with a horizontal offset) is presented alone, the direction of the horizontal offset is easy to report. This task becomes harder in the presence of a surrounding square flanker (Fig. 1b, column 1). Interestingly, the *global* configuration of flankers across the entire visual field determines crowding. For example, adding flankers as far away as 8.5 degrees from the 200 arcsec target can *improve* performance depending on the global configuration (*uncrowding*; Fig. 1b; Manassi, Lonchampt, Clarke, & Herzog, 2016; Manassi, Sayim, & Herzog, 2012). This strong dependency of performance on global configurations provides a qualitative signature which can easily be tested in models. Importantly, (un)crowding occurs across multiple paradigms (Herzog & Fahle, 2002; Pachai, Doerig, & Herzog, 2016; Sayim et al., 2010) and is not restricted to vision (Oberfeld & Stahn, 2012; Overvliet & Sayim, 2016). Hence, (un)crowding is not an idiosyncratic effect related to a specific paradigm. It rather reflects a general strategy used by the brain. This kind of general strategy for vision is precisely what we expect models to explain.

Crowding effects have been shown in ffCNNs (Doerig, Bornet, et al., 2019; Lonnqvist, Clarke, & Chakravarthi, 2019; Volokitin, Roig, & Poggio, 2017), and may occur by pooling the target and nearby flankers along the processing hierarchy. We hypothesize that this mechanism may not produce uncrowding because simple pooling can only deteriorate target-relevant information when flankers are added (Fig. 1c). However, intuitions are not to be trusted in complex systems with millions of parameters. Furthermore, new global processing strategies may emerge in shape-biased networks such as Geirhos et al.'s. Hence, it is currently unclear whether ffCNNs can carry out human-like global computations that lead to (un)crowding. Here, we thoroughly investigated (un)crowding in AlexNet (Krizhevsky et al., 2012), an ffCNN that was used as a model of the human visual system (Khaligh-Razavi & Kriegeskorte, 2014; Zeiler & Fergus, 2014), ResNet50 (He et al., 2016), a more sophisticated ffCNN, and the shape-biased network by Geirhos et al. (2018). We provide experimental evidence suggesting that it is the *architecture* of ffCNNs that prevent them from performing human-like global computations, and not the training procedure.

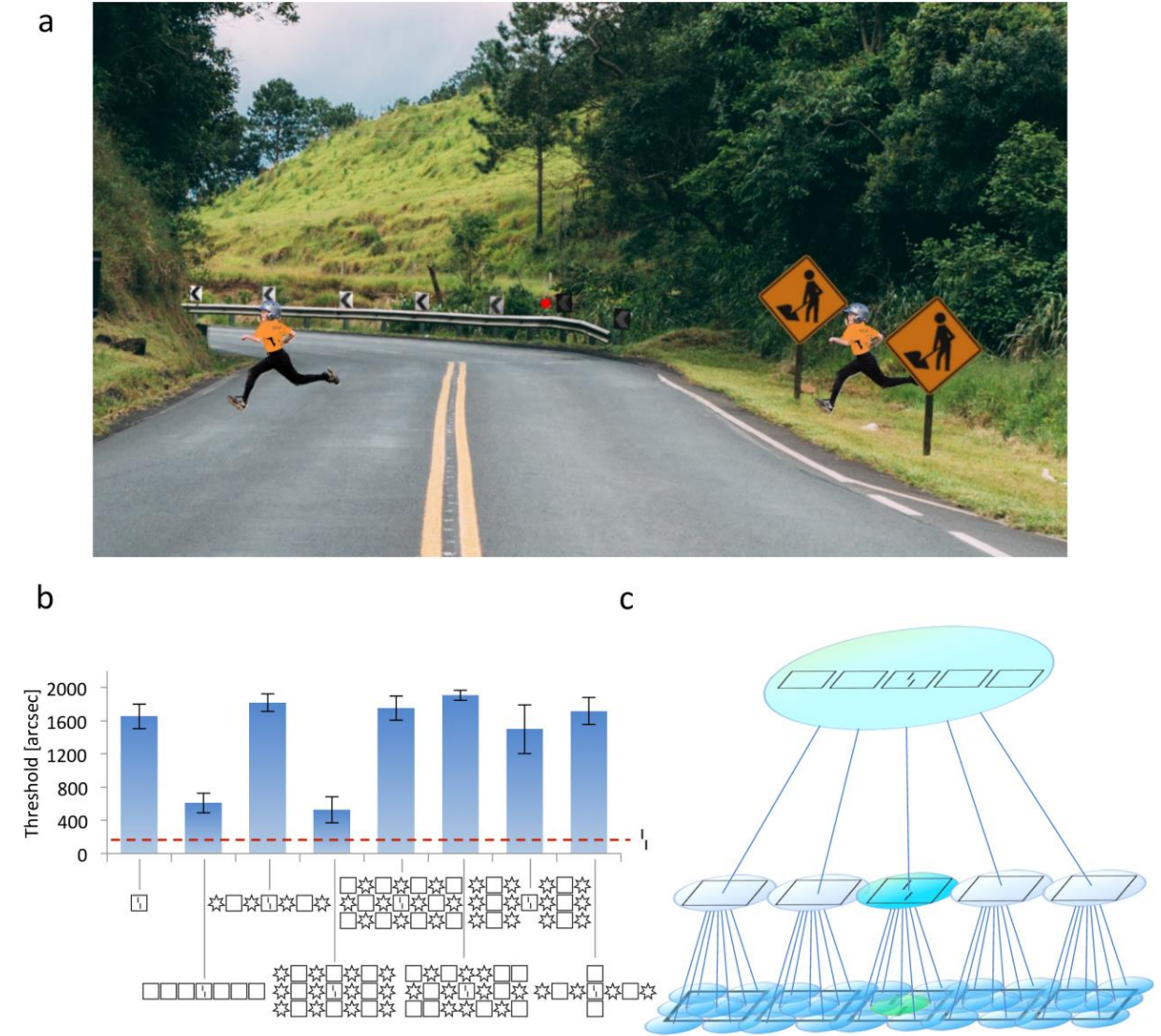

**Figure 1. Crowding. a. Crowding in a natural scene.** When fixating on the central red dot, it is more difficult to spot the kid on the right than on the left, because of the nearby signposts. Figure reproduced from Doerig, Bornet, et al. (2019) **b. (Un)crowding:** Manassi et al. (2016) presented a vernier either alone (red dashed line) or surrounded by a flanker configuration (x-axis). The y-axis shows the offset for which observers correctly report the vernier offset direction in 75% of the trials (threshold; performance is good when the threshold is low). When the vernier is presented alone, the task is easy (red dashed line). Adding a flanking square (column 1) makes the task much harder, a classic crowding effect. When more squares are added, performance recovers almost to the unflanked level (second

column, *un*crowding). Uncrowding strongly depends on the configuration (columns 2 to 8). For example, column 4 shows a configuration of flankers with a strong uncrowding effect. In comparison, column 5 has the same flankers but in a different configuration producing strong crowding. Modified from Doerig, Bornet, et al. (2019). **c. Crowding in ffCNNs:** In the feedforward framework of vision, embodied by ffCNNs, crowding occurs by pooling of visual features across a hiererachy of local feature detectors. In this example, a stimulus with five squares and a vernier target is presented. Each circle represents a neuron and shows the elements in its receptive field. In early layers, receptive fields are small and the vernier is in the receptive field of a single neuron (green). Neighboring neurons respond to parts of the squares (blue). At this level, the vernier is well represented. In the next layer, however, information about the vernier is pooled with information of the sourrouding flanker. Vernier-related information is "corrupted" by the flankers, making the offset direction harder to decode (crowding; blue-green). In subsequent layers, even more target-unrelated information is pooled. For this reason, we hypothesize that adding more flankers may always lead to more crowding in ffCNNs.

## Methods

Code and supplementary material are available online at https://github.com/adriendoerig/Doerig-Bornet-Choung-Herzog-2019.

### Experiment 1a

We presented different (un)crowding stimuli to AlexNet (trained on ImageNet prior to our experiment) and assessed how information about the target vernier is preserved along the network hierarchy. We used decoders to detect vernier offset direction based on the activity in each layer (Fig. 2). Each layer had its own decoder, consisting of batch normalization (Ioffe & Szegedy, 2015), followed by a hidden layer of 512 units, followed by an ELU non-linearity (Clevert, Unterthiner, & Hochreiter, 2015), finally projecting to a softmax layer composed of 2 nodes coding for left and right offsets. The weights of AlexNet were frozen during this process, only the decoder weights were trained. The decoders were trained using Adam optimizers (Kingma & Ba, 2014) to minimize the cross-entropy between the predicted and the presented vernier offsets. Each image in the training set consisted of a vernier plus a non-overlapping random configuration of flankers (composed of 18x18 pixels squares, circles, hexagons, octagons, stars or diamonds). These configurations had between 1 and 7 columns and between 1 and 3 rows of flankers of the same shape. We added Gaussian noise to each image. Training was successful, i.e., the network was well able to detect the vernier offset direction in the training images.

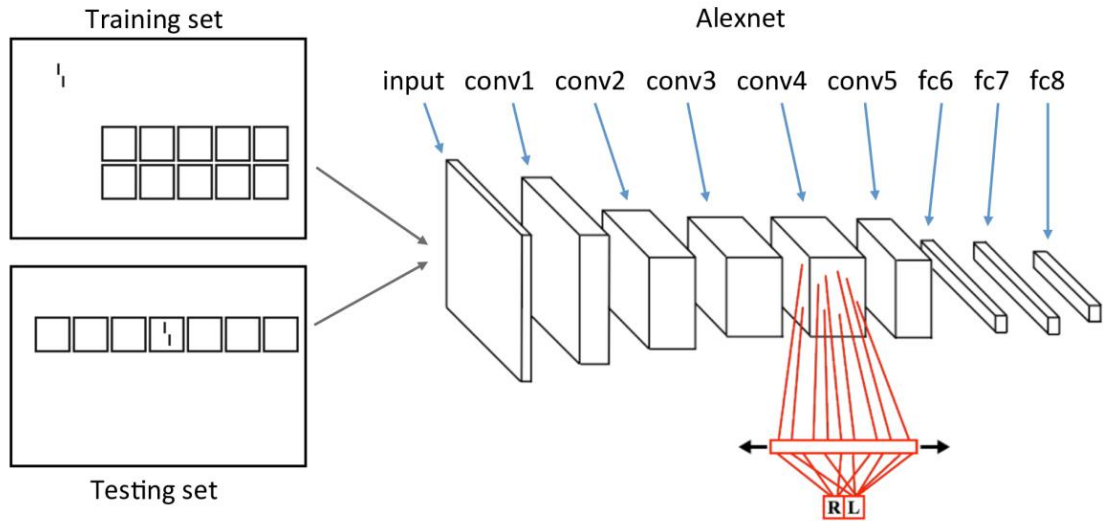

**Figure 2. Measuring crowding in ffCNNs.** To investigate how well information about the vernier offset is preserved throughout the network hierarchy, we trained one decoder (in red) at each layer to discriminate the vernier offset direction based on the activity elicited by the stimulus in this layer. For example, the stimulus at the top left of this figure is presented. This elicits activities in each layer of AlexNet and the decoders are trained to retrieve the offset direction based on this activity. Only the decoders are trained (red). In the training set, the vernier and a flanker configuration were simulatneously shown, but never overlapped (top). In the testing set, we presented 72 different (un)crowding configurations and measured performance for each configuration and each layer. In these testing images, the vernier was always surrounded by the flanker configuration (bottom). In this example, configurations of squares are shown, but we also used different shapes (see main text).

Our main question was how the network generalizes to the (un)crowding stimuli. Importantly, during training, the vernier target and the flanking configurations were presented simultaneously but never overlapped (Fig. 2). During testing the vernier was surrounded by different flanker configurations, as in the psychophysical (un)crowding stimuli (Fig. 2). The testing set consisted of 72 different configurations of flankers with Gaussian noise. There were 6400 trials per configuration with the configuration presented at different locations. For each layer of AlexNet, performance was measured as the proportions of correct vernier offset discrimination made by the decoder. We repeated this entire procedure 5 times, including training and testing, and report averaged performances.

Experiment 1b

We tested an ffCNN with a more sophisticated architecture (ResNet50) trained on ImageNet, and the same ffCNN architecture trained on a dataset tailored to bias the network towards global shape computations (i.e., Geirhos et al.'s shape-biased version of ResNet50). To this end, we applied exactly the same procedure as in experiment 1a to both the original version of ResNet50 and Geirhos et al.'s shape-biased version. The only difference was that we used 64 hidden units instead of 512, because this achieved better performance (i.e., better classification performance on crowded conditions).

Experiment 2

In experiment 2, we investigated which parts of the stimulus configurations the network mainly relies on by using an occlusion sensitivity measure (similarly to Zeiler & Fergus, 2014). We used the networks with decoders trained in experiment 1. For a given configuration, we collected the vernier offset decoder's output at each layer. Then we slid a 6x6 pixels Gaussian noise patch over the entire configuration and measured for each patch position P and network layer L how much the noise patch affected the vernier offset discrimination. The noise patch had the same statistics as the background noise, effectively removing parts of the stimulus. The rationale is that when the patch occludes parts of the stimulus, which are important for classification, decoder predictions should be strongly affected. On the other hand, if the patch occludes an unimportant part of the stimulus, decoder predictions should not be affected. Since the global stimulus configuration matters for uncrowding, we were interested to see if the network relies on the global configuration or if it simply focused on the region close to the vernier.

For each patch location P and layer L, we quantified how much the noise patch biased vernier offset classification towards or away from the correct response:

$$score_{P,L} = \frac{\{\vec{T} \cdot (\overrightarrow{y_{P,L}} - \overrightarrow{x_L})\}_{left\_vernier}}{2} + \frac{\{\vec{T} \cdot (\overrightarrow{y_{P,L}} - \overrightarrow{x_L})\}_{right\_vernier}}{2}$$

Where $\vec{x_L} = (x_1, x_2)_L$ is the output of the decoder for layer L on the original stimulus *without* a noise patch ($x_1$ and $x_2$ respectively correspond to the network's prediction for a left- or right-offset vernier), $\vec{y_{P,L}} = (y_1, y_2)_{P,L}$ is the output of the decoder for layer L *with* the noise patch at position P and $\vec{T}$ is a vector equal to $(+1, -1)$ if the correct vernier offset is left and $(-1, +1)$ otherwise. To avoid biases related to offset direction, we computed the mean score of the left- and right-offset versions of each stimulus.

Using this procedure, we obtained maps indicating which regions of a stimulus are most important for vernier offset discrimination. We used four different stimuli from Manassi et al. (2016): a vernier alone, a vernier flanked by one square (leading to crowding in humans), a vernier flanked by a row of seven squares (leading to uncrowding in humans), and a vernier flanked by a row of seven alternating squares and stars (no uncrowding in humans). Additional stimuli are shown in the supplementary material.

# Results

### Experiment 1a

Unlike humans, AlexNet shows crowding but *not un*crowding. The vernier offset is easily decoded from each layer when the vernier is presented alone, and performance drops when a single flanker is added. Crucially, performance deteriorates further when more flankers are added, regardless of the shape type (Fig. 3a). Squares produced more crowding than circles, hexagons, octagons or diamonds, presumably because the vertical bars of the squares interfered with the vernier more strongly. These results hold for all layers of AlexNet (supplementary material).

Fig. 3b shows that, unlike humans who show strong uncrowding depending on the configuration, only the number of shapes seems to affect crowding in AlexNet – and not the configuration. Although certain configurations with three flankers have a higher percentage of correct response than certain configurations with a single flanker, this effect is driven by the shape type and not by the configuration of shapes. For example, the networks are better at dealing with diamonds than squares (Fig. 3a; probably because squares interfere more with verniers due to the vertical orientation of their edges). Still, adding extra shapes always deteriorates performance compared to a single shape, regardless of the configuration. This pattern of results is similar in all layers of AlexNet (supplementary material).

**Experiment 1b**

We applied the same analysis to the original ResNet50 and Geirhos et al.'s shape-biased version of ResNet50. The results for both networks are qualitatively similar to the results for AlexNet in experiment 1a (Fig. 3c&d). One difference is that the performance of the decoder is always below chance level with diamonds. This indicates that information about the vernier offset survives, even though the diamond flanker reverses the prediction. Adding additional diamond flankers brings performance closer to chance level, indicating that less information about the vernier offset survives, i.e., crowding increases when adding flankers. Another difference is that the squares lead to the least amount of crowding, contrary to AlexNet.

First, these results show that using a more sophisticated ffCNN (i.e., ResNet50) does not allow ffCNNs to explain global uncrowding effects. Second, crucially, Geirhos et al.'s training method to bias ffCNNs towards shape does not lead to uncrowding either. This suggests that ffCNNs do not carry out human-like shape level computations for *architectural* reasons, and not because of the way they are *trained*.

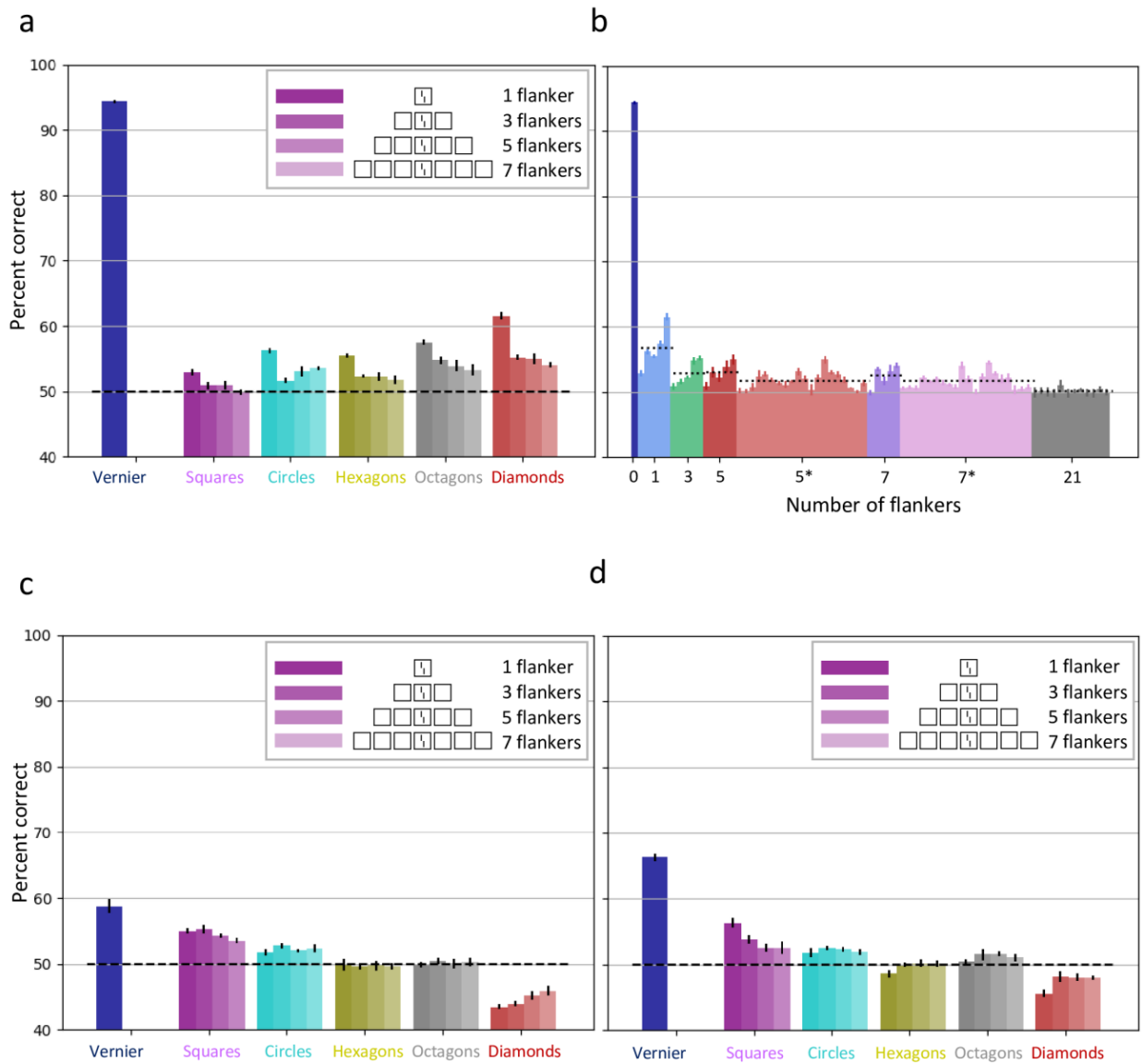

Figure 3. **a. Vernier offset discrimination performance for AlexNet with an increasing number of identical flankers.** The x-axis shows different flanker configurations. Each color corresponds to one flanker shape, and brighter colors indicate more flankers (from darkest to lightest: 1, 3, 5 & 7 identical flankers). The single dark blue bar on the left corresponds to the vernier alone condition. The y-axis indicates the percentage of correct vernier offset responses. Error bars show standard deviation (N=5). Unlike humans, for whom performance improves when more identical flankers are added (Fig. 1b, columns 1&2; Manassi et al., 2016), performance deteriorates or stagnates for AlexNet with all flanker shapes. The results of this figure are decoded from layer 5 of AlexNet. Decoding vernier offsets from the other layers in AlexNet led to similar results (see supplementary material). **b. Vernier offset discrimination performance for AlexNet with 72 configurations.** The x-axis shows different flanker configurations sorted by number of flankers. Different colors

correspond to different kinds of flanker configurations. The labels correspond to the number of flankers in the configuration, and an asterisk indicates alternating shapes (e.g. square-circle-square-circle-square). From left to right: vernier alone, single flanker, 3 identical flankers, 5 identical flankers, 5 flankers alternating between two shapes, 7 identical flankers, 7 flankers alternating between two shapes and configurations of 3x7 flankers. The y-axis indicates percent correct of vernier offset discrimination for each flanker configuration (the dashed lines shows the mean percent correct for each kind of flanker configuration). The results of this figure are decoded from layer 5 of AlexNet. Decoding vernier offsets from the other layers in AlexNet led to similar results (see supplementary material). **c&d. Vernier offset discrimination performance for (shape-biased) ResNet50 with an increasing number of identical flankers. c.** Original ResNet50 (trained on ImageNet). **d.** Geirhos et al.'s shape-biased version. The results for both of these networks are qualitatively similar for the AlexNet results in panel a. The results of this figure are decoded from the output of the third bottleneck unit (see our shared code and He et al., 2016). Decoding vernier offsets from the other layers led to similar results (see supplementary material).

**Experiment 2**

Uncrowding requires global computations across large regions of the visual space. The configuration in its entirety determines performance and not only the elements in the neighborhood of the target (Doerig, Bornet, et al., 2019; Manassi et al., 2016, 2012). As mentioned, it has been proposed that ffCNNs focus largely on local features. This is indeed what we observed in experiment 2 in AlexNet (Fig. 4), ResNet50 (supplementary material), and Geirhos et al.'s shape-biased version of ResNet50 (Fig. 4): only elements in a local region around the target matter for classification. The same results also hold for the eight other stimulus types we tested (supplementary material). In general, as expected, occluding the vernier target deteriorates performance and occluding parts of the flanker surrounding the vernier improves performance. Occluding other parts of the stimulus, however, does not generally affect performance. Certain cases are harder to explain, such as the 1square condition shown in the top right panel of Fig. 4, in which occluding parts of the vernier improved classification. Although we cannot provide a definitive explanation, we suggest that this may be due to the classifier confusing a vertical bar of the square with a vertical vernier bar. Alternatively, this may be due to the background noise present in each stimulus. In rare cases, the occluder has an effect even when it does not cover the stimulus (e.g. in the bottom right panel of Fig.4). These cases are also probably due to background noise. Aside from these small peculiarities, the finding that only elements in the neighborhood of the vernier affect classification is very stable over all stimuli and network layers (see images and animations in the supplementary material).

These results suggest that the inability of ffCNNs to explain uncrowding stems from their focus only on local features close to the vernier. Importantly, although Geirhos et al.'s shape-biased network is biased towards global features, still, performance seems determined only by elements close to the vernier.

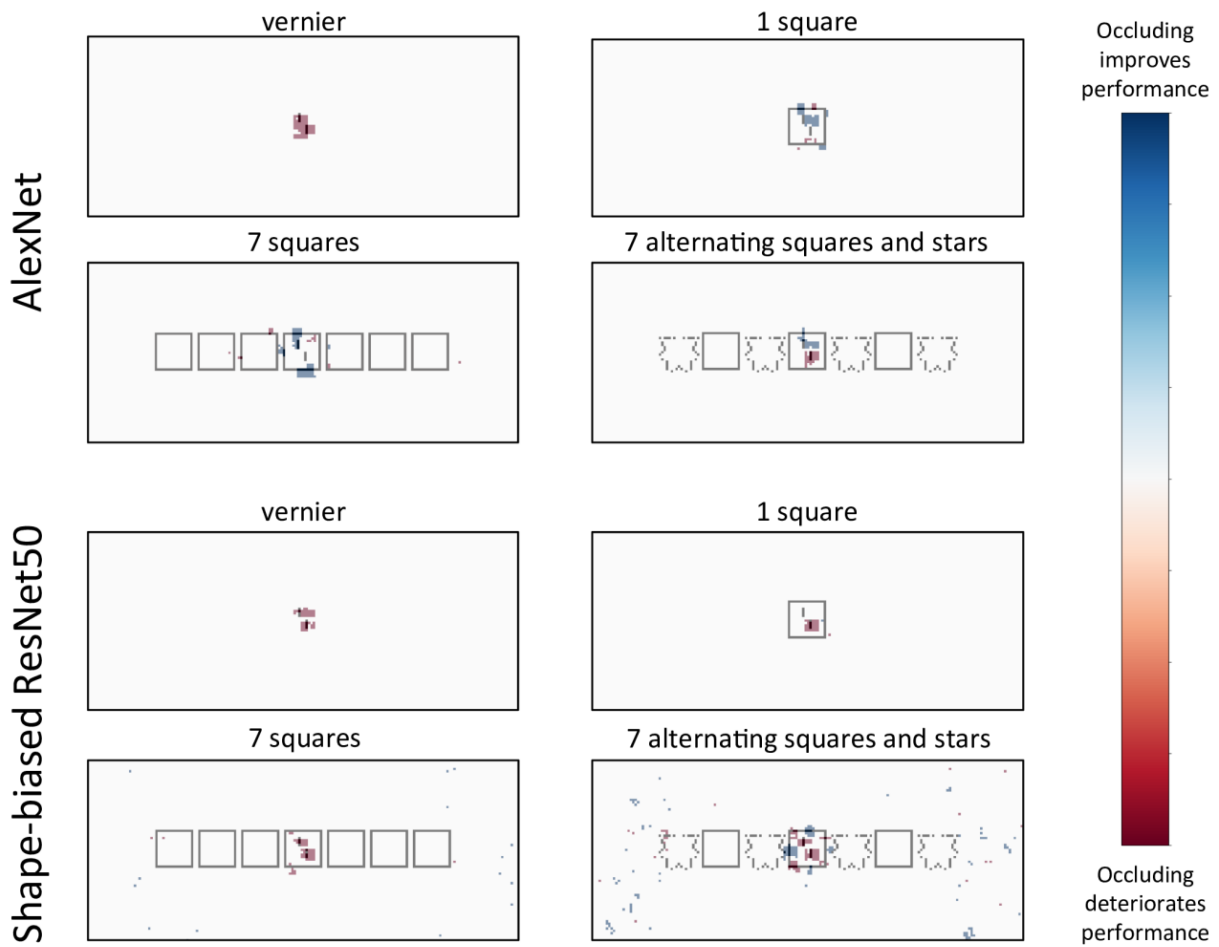

Figure 4: Occlusion analysis. Results of the occlusion analysis for AlexNet (*top*) and the shape-biased ResNet50 (*bottom*). Stimuli on the left lead to good performance in humans, while stimuli on the right lead to strong crowding in humans (Manassi et al., 2016). For both AlexNet and the shape-biased ResNet50, the network's decisions rely only on local elements in the target neighborhood regardless of the global stimulus configurations. We summed the maps for each layer to show which stimulus regions are most relevant across the network. We used all layers for AlexNet,

and, for the shape-biased ResNet50, the third convolutional layer in the first bottleneck plus the output of the first 9 bottleneck units (see our shared code and He et al., 2016). We then applied a threshold to each map at 0.4 times the maximal value in the map, for visibility. Per-layer results without thresholding can be found in the supplementary material, as well as animations showing what happens as the threshold value is changed. Results for the original ResNet50 and other layers of the shape-biased network are also shown in the supplementary material.

## Discussion

(Un)crowding is ubiquitous. It occurs in vision, audition and haptics (Manassi et al., 2016; Oberfeld & Stahn, 2012; Overvliet & Sayim, 2016; Whitney & Levi, 2011). This pervasiveness is not surprising because elements rarely appear in isolation. Any perceptual system needs to cope with crowding to process information in cluttered environments. (Un)crowding is a probe into how the visual system computes global information.

In this contribution, we asked whether large ffCNNs trained on complex visual tasks can explain (un)crowding. We chose this approach because these ffCNNs are often used as brain models. The idea is that the weights learned by these ffCNNs to solve complex visual tasks may lead to human-like visual processing. For this reason, we did not change the ffCNN weights for quantifying (un)crowding, i.e., we only trained the additional decoders. We found that these ffCNNs do not seem to carry out human-like global computations.

Experiment 1 shows that current ffCNNs do not explain (un)crowding. In other words, training an ffCNN on a complex natural image recognition task does not automatically yield a network performing similarly to the human visual system. Experiment 2 suggests that this is due to the inability of ffCNNs to take the entire stimulus configuration into account. In ffCNNs, only elements in the target's neighborhood affect performance. Global features do not affect how local parts are processed. In humans, on the other hand, the global configuration strongly affects processing of local parts. For example, vernier offset information can be "rescued" by certain global configurations.

This difference could not be remedied by a different *training* protocol. Indeed, all our results also hold for Geirhos et al.'s shape-biased ffCNN. We suggest that, although Geirhos et al.'s training procedure successfully biased the networks towards global features, it does not show human-like global shape computations. Indeed, the network still seems limited to combining features by pooling along the feedforward cascade. Hence, unlike in humans, global configuration cannot affect processing of local parts. For these reasons, our results suggest that the inability of ffCNNs to perform human-like object shape processing is rooted in their feedforward pooling *architecture*. Because of this pooling, performance deteriorates when flankers are added. For this principled reason, we propose that ffCNNs cannot produce uncrowding in general, independently of the specific ffCNN, training procedure and loss function. In support of this proposal, we showed in a separate contribution that ffCNNs specifically trained on classifying verniers and flanking shapes, as well as counting the number of flankers, do not produce global (un)crowding either (Doerig, Schmittwilken, Sayim, Manassi, & Herzog, 2019).

Global processing is not only an issue for ffCNNs but for other models too. We showed that no existing model of crowding based on local and feedforward computations can explain uncrowding (Doerig, Bornet, et al., 2019; Herzog & Manassi, 2015; Manassi et al., 2016; Pachai et al., 2016). There seems to be a principled difference in computational strategies, based on architecture, between humans and feedforward pooling systems.

Hence, despite their well-known power, further aspects need to be incorporated into ffCNNs. We propose that recurrent, global grouping and segmentation is crucial to explain how the brain deals with global configurations (Doerig, Bornet, et al., 2019; Doerig, Schmittwilken, et al., 2019). Specifically, we propose that a flexible recurrent grouping process determines which elements are grouped into an object. In the case of (un)crowding, elements are first grouped together and then only elements within a group interfere with each other. If the configuration of flankers ungroups from the target, the target is released from crowding. Francis, Manassi, and Herzog (2017) proposed a spiking neural network with a dedicated recurrent grouping process, which is able to explain why (un)crowding occurs (see also Bornet et al., 2019). However, this model is tailored to group oriented edges and cannot generalize to grouping of more complex features. Deep learning models are promising because they are more flexible and can be trained to deal with any kind of stimulus.

Doerig, Schmittwilken, et al. (2019) showed that capsules networks (Sabour, Frosst, & Hinton, 2017), combining CNNs with a recurrent grouping and segmentation process, can explain (un)crowding, including temporal characteristics of uncrowding. Linsley et al. (2018) proposed recurrent grouping and segmentation modules to improve CNNs, and there are several other approaches to experiment with grouping and segmentation in recurrent network architectures (Lotter, Kreiman, & Cox, 2016; Nayebi et al., 2018; Spoerer, Kietzmann, & Kriegeskorte, 2019; Spoerer, McClure, & Kriegeskorte, 2017). More work is needed to compare and characterize computations in different recurrent architectures.

Our results contribute to the expanding literature showing that there is much more to vision than combining local feature detectors in a feedforward hierarchical manner (Baker et al., 2018; Brendel & Bethge, 2019; Doerig, Bornet, et al., 2019; Doerig, Schmittwilken, et al., 2019; Funke et al., 2018; Kar, Kubilius, Schmidt, Issa, & DiCarlo, 2019; Kietzmann et al., 2019; Kim, Linsley, Thakkar, & Serre, 2019; Lamme & Roelfsema, 2000; Linsley et al., 2018; Sabour et al., 2017; Spoerer et al., 2019, 2017; Tang et al., 2018; Wallis et al., 2019). In line with the present findings, many studies have highlighted other fundamental differences between ffCNNs and humans in local vs. global processing. For example, Baker et al. (2018) showed that ffCNNs but not humans are affected by local changes to edges and textures of objects. Brendel and Bethge (2019) showed that ffCNNs classify ImageNet images almost as well when using small local image patches than when using the entire images. These results clearly show that image classification is underconstrained as a testbed. For this reason, well-controlled psychophysical stimuli, which allow detailed analysis, should be used in addition to image classification (RichardWebster, Anthony, & Scheirer, 2018). Simply testing whether deep learning systems reproduce idiosyncratic illusions, without linking them to computational mechanisms, does not provide principled insights. Hence, an important question will be what are the crucial benchmarks targeting principled computational processes. Here, using crowding, we showed a fundamental difference in local vs. global processing between humans and ffCNNs, and suggest that grouping and segmentation are promising additions to make deep neural networks better models of vision.

Historically, psychophysical results were seen as stepping stones towards object recognition models. Today, the picture has been reversed: we have powerful artificial vision models, but they do not reproduce even simple psychophysical results. The fact that ffCNNs can solve complex visual tasks in a different way than humans reveals that there are many ways of doing so. There are many roads to Rome. Despite the diversity of possible strategies to solve complex vision tasks, deep insights can be derived by comparing the crucial underlying computations adopted by different systems.

# Acknowledgements

Adrien Doerig and Oh-Hyeon Choung were supported by the Swiss National Science Foundation grant n.176153 "Basics of visual processing: from elements to figures". Alban Bornet was supported by the European Union's Horizon 2020 Framework Programme for Research and Innovation under the Specific Grant Agreement No. 785907 (Human Brain Project SGA2). The funders had no role in study design, data collection and analysis, decision to publish, or preparation of the manuscript.